# Probabilistic Resilience of DER Systems – A Simulation Assisted Optimization Approach

Sakshi Mishra
Energy Systems Integration Department
National Renewable Energy Laboratory
Golden, United States
Sakshi.m@outlook.com

Kate Anderson
Energy Systems Integration Department
National Renewable Energy Laboratory
Golden, United States
Kate.Anderson@nrel.gov

*Abstract*— Energy systems resilience is becoming increasingly important as the frequency of major grid outages increases. In this work, we present a methodology to optimize a behind-the-meter distributed energy resource system to sustain a site's critical loads during a pre-defined outage period. With the fixed system design, we then propose an outage simulation approach to estimate the resilience potential of the DER system to sustain loads beyond the fixed outage period – a yearlong resilience performance analysis. We apply statistical analysis to assess the system's resilience performance over a broader parametric problem space on an hourly, monthly, and yearly basis. We demonstrate the impact of the pre-defined outage period on the resilience performance through a case study. Results show that the probability of surviving a random outage of a given duration changes from 20% to 95% when the outage is modeled for a weekday instead of a weekend for the given load-profile.

## I. Introduction and Background

Because critical infrastructure sectors such as healthcare, water, defense, and communications rely on electricity, the energy sector has been deemed uniquely critical infrastructure by the Presidential Policy Directive 21 (PPD-21) [1]. The loss of electricity disrupts day-to-day critical operations of a community, resulting in health and safety impacts as well as economic losses [2]. As the frequency of natural and man-made disasters and corresponding grid outages increases [3], building resilience in energy infrastructure is rapidly becoming a national priority for countries around the world, including the U.S. [2].

A resilient power system has been defined as "a grid which has four fundamental properties of resilience, namely anticipation, absorption, recovery, and adaptability after the damaging events" [4]. Increasing resilience requires the strategic hardening of both the physical and cyber components of the power system [4] [5]. Hardening utility-scale generation, transmission, and distribution infrastructure requires significant capital investment and can face complex regulatory and policy hurdles, making it a relatively slow process. Localized Behind-the-Meter (BTM) resilience solutions, on the other hand, can be built and deployed faster, with fewer regulatory hurdles. These solutions often include Distributed Energy Resources (DERs) which can also provide economic benefits to the site owner.

The work presented in this paper focuses on assessing the resilience potential of BTM DER systems. We define resilience performance as the amount of time a BTM DER system can sustain the critical load without utility power supply, for given system design, load profile, and geographic location. We simulate outages starting every hour of the year and evaluate how survival varies across the year. We find that the ability of systems to sustain critical loads is highly dependent on the load profile, resource availability, and the hour of the year when the outage starts, reflecting weekly and seasonal variations.

## II. REopt Lite – A Tool To Design Resilient Systems

In this work, we employ REopt Lite, a Mixed Integer Linear Programming (MILP) based optimization formulation to determine the optimal DER system sizes and dispatch strategy. The optimum system minimizes the lifecycle cost of energy to the site while ensuring the critical load is sustained without the utility grid during a predefined outage period. We call this step *optimization module* (OM). The second step, which is the focus of this paper, then assesses the overall resilience performance of the system during other potential outages throughout the year. We call this step *simulation module* (SM), as shown in Figure 1.

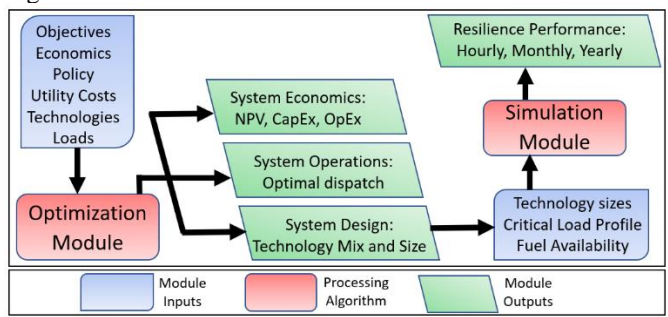

Figure 1. Sequential Processing of Optimization and Simulation Modules

The following two sub-sections further elucidate the functionalities of these two modules. It is important to note that the novel contribution of this paper is the SM methodology – its in-depth description and analysis are presented in sections III, IV, and V.

### A. The optimization module

The OM takes in the load profile, utility rate tariff, economic parameters (technology capital costs, fixed and variable operations and maintenance costs, escalation and discount rates), policy information (incentives and net-metering), and technology options (solar, wind, combined heat and power [CHP], storage, utility grid, and any existing on-site DER). The deterministic MILP formulation recommends the optimal technology mix, size, and dispatch, and the resulting project lifecycle economics. The objective function in the mathematical model is formulated for cost minimization – it considers capital costs of technologies, fixed and variable O&M costs, fuel charges, grid energy & demand charges, production incentives as well as energy export payments. Constraints include physical operational limits of generation technologies and storage, load balance, rate tariff (net-metering limits), etc. [6] has the full mathematical formulation documented. Further details are presented in [7] and [8].

The OM is designed to conduct two types of analysis – i) Financial and ii) Resilience. In the financial analysis, REopt Lite finds the system design and dispatch that minimizes the lifecycle cost of energy to the site. The resilience analysis does the same thing, but with the added constraint that the system must sustain the critical load (often some fraction of the typical load) without the utility grid during the specified outage period.

### B. Simulation Module

The SM accepts the OM-recommended technology type and their respective sizes as inputs. It then assesses the resilience performance of the system for a full year by simulating outages starting every hour of the year (8760 times). The next section discusses the SM in detail.

## III. SIMULATION MODULE

The major difference between the OM and the SM is the modeling methodology. The OM fixes the outage period and optimizes system sizes and dispatch strategy. The SM, on the other hand, takes in fixed system sizes and simulates outages starting at each hour of the year instead of a single outage period.

The SM dispatches the on-site assets to meet the critical load on an hourly or sub-hourly basis using a load-following strategy. When the technology mix includes a conventional generator, each outage simulation starts with a fixed quantity of on-site fuel defined by the user, and we assume fuel cannot be re-supplied during a given outage (as commonly experienced during disasters). When the technology mix includes a battery, the SM obtains the battery state-of-charge in each hour from the OM, which is typically between 20% and 100%. The state-of-charge depends on how the battery is used for peak management and energy arbitrage while grid-connected.

For each of the 8760 simulations, the technologies are dispatched to meet the critical load until there are not enough resources to meet the load in a given hour, or until the simulation reaches a full year (8760 hours). The SM calculates the length of the survived outage durations for outages starting every hour of the year and then evaluates the resilience performance of the system. The load following strategy and the probabilistic resilience performance evaluation are described in more detail in the following sections. The code for the OM and SM is available on GitHub [9].

### A. Load Following Dispatch Strategy

The load following strategy is used to determine the hourly dispatch in each outage simulation. It is implemented as follows:

- Subtract solar and wind generation from the critical load.

- If the combined solar and wind generation is more than the critical load, and there is space available in the battery, charge the battery with the surplus generation. Or else, let the surplus generation dissipate.

- If the critical load is still unmet (after subtracting solar and wind generation), check if the diesel generator can supply the remaining critical load by verifying the generator capacity is greater than or equal to the remaining load and there is sufficient fuel for the generator to run. Fuel availability calculated using the diesel fuel burn rate.

- If the generator has a minimum turndown limit that requires generating more than the unmet critical load, then charge the battery with the excess energy. If the battery's state of the charge is at its maximum, then the energy is dissipated using the 'dump load resistor'.

- If the critical load is still unmet, discharge the battery within its capacity and minimum state of charge constraints.

- If the critical load is still unmet, break the simulation loop for the outage starting in that specific hour.

### B. Probabilistic Resilience Performance Evaluation

The survived outage durations for outages starting every hour of the year is an array of length 8760 for hourly analysis, denoted by r, where each value in the array is the number of hours survived by the system for the outage starting in the [index +1]$^{th}$ hour. After calculating the survived outage durations (or r) series, the probabilities of survival for outages of various durations are calculated using the following formula:

$$P(hrs_i) = \frac{1}{ts} \sum_{h \in r} \begin{Bmatrix} 1 & if\ h > hrs_i \\ 0 & otherwise \end{Bmatrix} \quad for\ i\ \in [1, r_{max}]$$

Note that the equation above is centered using a center tab stop. Where $P(hrs_i)$ the probability of is surviving $i$ number of hours; $ts$ is the number of time steps (8760 for hourly analysis); $h$ is the number of hours survived for an outage starting in the $i^{th}$ hour (which is a value from the $r$ series); $hrs_i$ is the number of hours survived of which the probability is being calculated; $r_{max}$ is the maximum number of hours the system survived (for an outage starting a particular hour of the year – given by the index of $r_{max}$ in the series). The above calculation provides the probabilities of surviving an outage of lengths ranging from 1 to $r_{max}$. These probabilities are averaged over hour-of-the-day (for all 24 hours) and month-of-the-year (for all 12 months) to capture the effect of outage start hour and start month on the outage survival length.

## IV. CASE STUDY

In this case study, we investigate how daily variations in the load profile of a site (during the OM-modelled outage period) impact the sizing of the recommended technologies and the resilience performance of the system. The case study site is in Palmdale, CA. It is a retail store building with an annual energy consumption of 1,000,000 kWh on a time-of-use utility tariff [10]. The load profile is simulated using the DoE commercial reference retail store building for this climate zone. The load profile, shown in Figure 2, shows clear weekday vs. weekend pattern differences. The pre-defined outage period is 24 hours. The critical load is 50% of the typical load. Also, Figure 2 illustrates the full-load profile, as opposed to the load profile shown in the subsequent figures where it is attenuated to 50% (critical load) during the outage period. The following two subsections discuss the results of two scenarios when the pre-defined outage is modeled on a weekday versus a weekend.

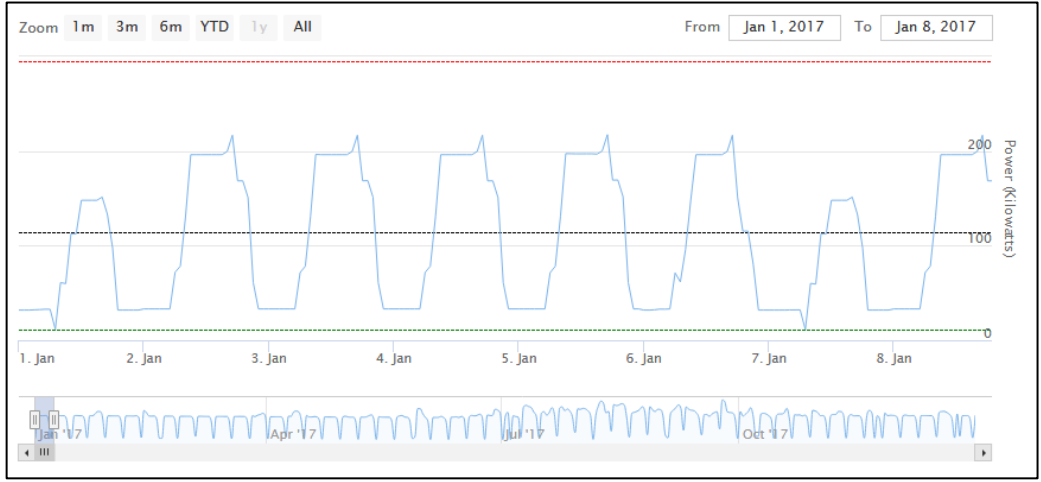

Figure 2. January Load Profile (Week 1)

### A. Scenario I - outage starting on a weekend

For this scenario, the pre-defined outage is modeled to start at midnight on January 1st, 2017, which is a Sunday. The OM recommends a 375 kW PV system, 82 kW; 282 kWh battery, and 5 kW diesel generator to provide energy to meet the typical energy needs of this site at lowest lifecycle cost, while also sustaining the specified outage. The dispatch plot shown in Figure 3 illustrates how PV, battery, and diesel combine to sustain the 50% critical load during the outage with the power (kW) on the left y-axis, battery state of charge on the right y-axis, and the date and time on the x-axis. After the outage, we see the grid recharge the battery and serve the load.

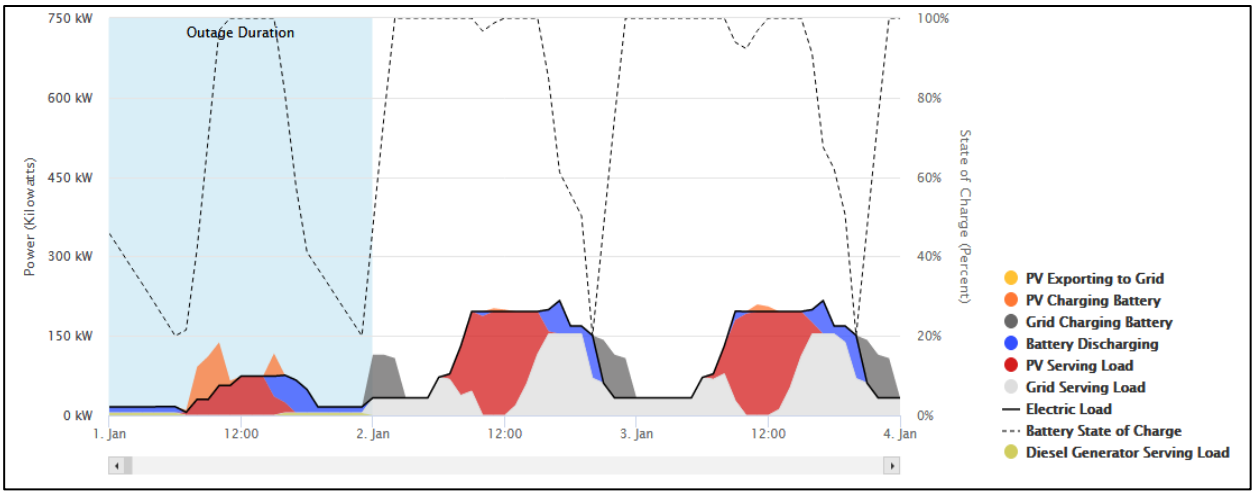

Figure 3. A 375 kW PV system, 82 kW; 282 kWh battery, and 5 kW diesel generator sustain the critical load during a weekend outage.

While the system design can sustain the specified outage on January 1st, the SM shows that the probability of sustaining a 24-hour long outage starting at any hour of the year, not just January 1st, is only 20%. Figure 4 shows the probabilities of surviving an x hours-long outage (where x is the x-axis value) for an outage that starts at the given hour of the day. We see that the system has a lower probability of sustaining outages that start in the late afternoon and evening, because there is little solar generation during these hours. The battery and diesel generator often cannot sustain the load on their own until solar generation is available again the next day.

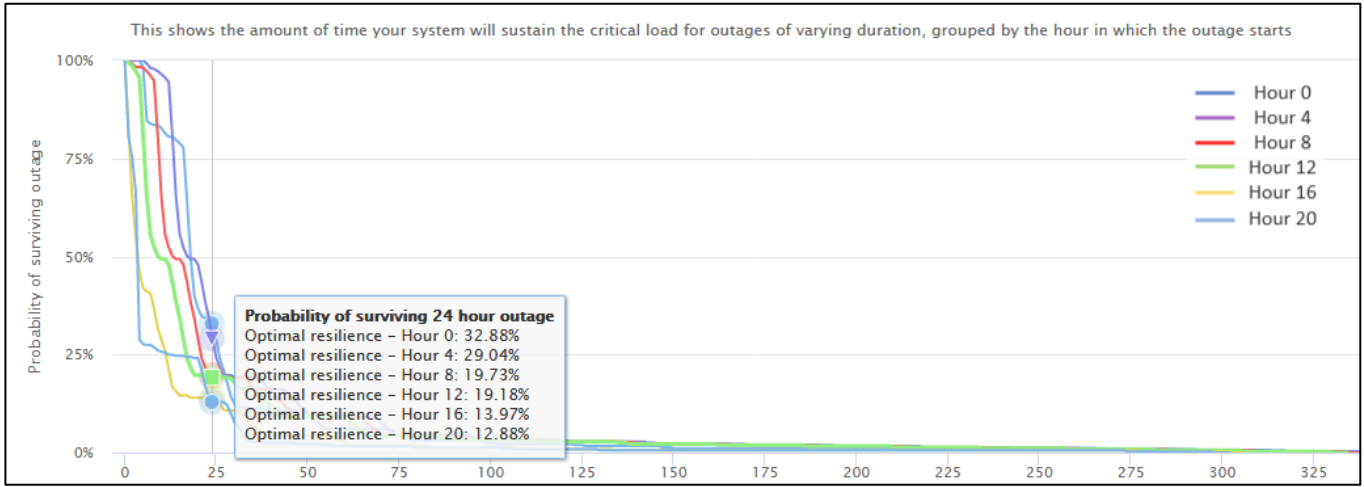

Figure 4. The system designed for a weekend outage has a lower probability of surviving other outages of similar length throughout the year

### B. Scenario II – outage starting on a weekday

Keeping all other inputs constant from scenario I, in scenario II, the pre-defined outage is modeled to start at midnight on January 2nd, 2017 (Monday). The OM recommends a 386 kW PV system, 92 kW; 375 kWh battery, and 33 kW diesel generator to meet the typical energy needs of this site at lowest lifecycle cost, while also sustaining the specified outage.

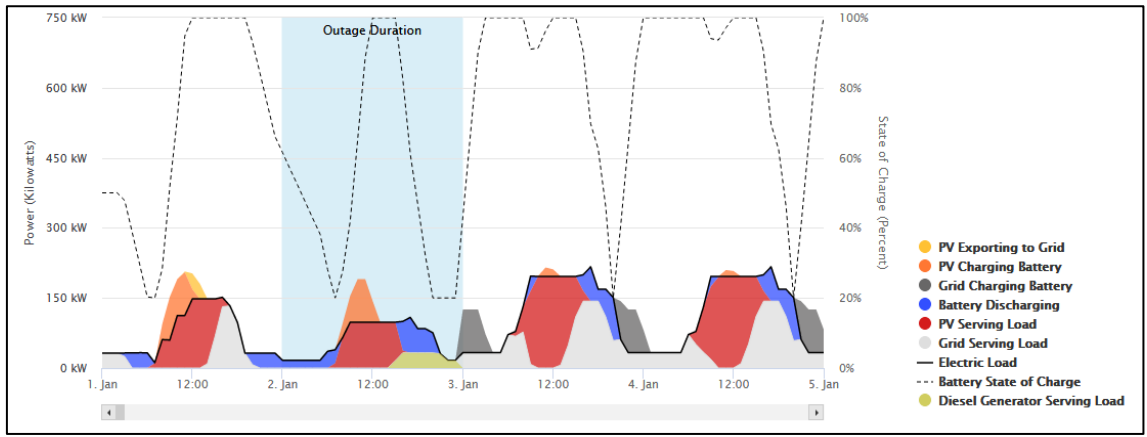

Figure 5. A larger PV system, battery, and diesel generator are required to sustain the higher critical load during a weekday outage

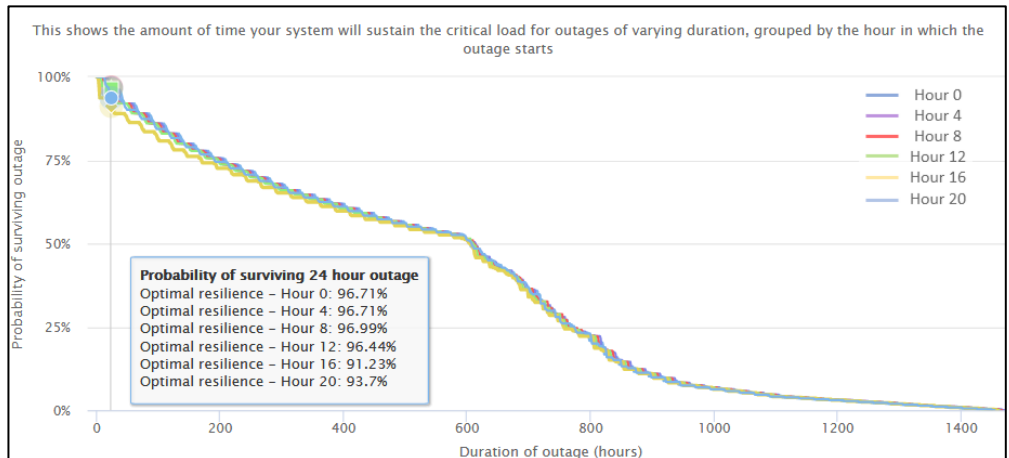

Figure 6. The system designed for a weekday outage in January has a higher probability of surviving longer duration outages throughout the year.

The dispatch plot for this scenario (Figure 5) shows the critical load (black line) on the weekday is higher than the weekend critical load modeled in Scenario I, resulting in larger required systems sizes (in particular, a much larger diesel generator). The SM calculates the probability of sustaining a 24-hour long outage starting at any hour of the year, not just January 2nd, is 95% with this system design. The survivability probabilities are drastically different in the two scenarios, as evident from the plots shown in Figure 4 and Figure 6. The maximum number of hours that can be survived by the system in Scenario I is ~340 hours and average is 21 hours (x-axis), whereas in Scenario II, maximum reaches ~1450 hours and average is 523 hours. This case study demonstrates that the outage period selected for the OM strongly impacts the optimal system size recommended; which in turn, dictates the resilience performance of the system simulated by the SM. While the economics of the two scenarios is out of scope for this paper, more information related to the economics can be retrieved from [11] and [12].

## V. SUMMARY AND OUTLOOK

In this work, we propose a probabilistic methodology to determine the resilience performance of a BTM DER system. An optimal system design configuration, capable of meeting the critical load for a pre-defined outage time and duration is determined by the OM. The resulting technology types and their respective sizes are then fed to the SM for assessing the resilience performance of the system probabilistically. When coupled with the techno-economic insights about the BTM DER project's viability, insights into the resilience performance of the system can further strengthen the case for DER systems. A functional version of this work is made available for free public use through the REopt Lite tool [13].

REopt Lite has recently added combined heat and power (CHP) to the suite of technologies in the OM. Future work will add CHP to the resilience assessment. One of the shortcomings of the present approach is its deterministic nature – the system is designed with a pre-defined outage. Applying a stochastic approach to modeling the outage time and duration in the OM is a relevant future research direction to pursue. Furthermore, the SM can be improved to consider the intermittency in the solar and wind outputs as another interesting future enhancement to the presented approach.

## ACKNOWLEDGMENT

This work was authored by the National Renewable Energy Laboratory, operated by Alliance for Sustainable Energy, LLC, for the U.S. Department of Energy (DOE) under Contract No DE-AC36-08GO28308. Funding provided by the U.S. Department of Energy Office of Energy Efficiency and Renewable Energy Federal Energy Management Program. The authors would like to thank Kai-Wei Yu for help with coding the outage simulation module; Emma Elgqvist and Linda Parkhill for reviewing and testing the model. The views expressed in the article do not necessarily represent the views of the DOE or the U.S. Government. The U.S. Government retains and the publisher, by accepting the article for publication, acknowledges that the U.S. Government retains a nonexclusive, paid-up, irrevocable, worldwide license to publish or reproduce the published form of this work or allow others to do so, for U.S. Government purposes.

## REFERENCES

[1] T. W. House, "Presidential Policy Directive -- Critical Infrastructure Security and Resilience," 2013.

[2] T. W. House, "Economic benefits of increasing electric grid resilience to weather outages," Executive Office of the President, 2013.

[3] DOE, "OE-417 Electric Emergency and Disturbance Report - Calendar Year 2019," 2019.

[4] F. H. Jufri, V. Widiputra and J. Jung, "State-of-the-art review on power grid resilience to extreme weather events: Definitions, frameworks, quantitative assessment methodologies, and enhancement strategies," *Applied Energy,* vol. 239, pp. 1049-1065, 2019.

[5] N. Bhusal, M. Abdelmalak, M. Kamruzzaman and M. Benidris, "Power System Resilience: Current Practices, Challenges, and Future Directions," *IEEE Access,* vol. 8, pp. 18064-18086, 2020.

[6] L. REopt, "Mathematical formulation (https://github.com/NREL/REopt_Lite_API/wiki/REopt-Mathematical-Model-Documentation)," 2020.

[7] S. Mishra, J. Pohl, N. Laws, D. Cutler, T. Kwasnik, W. Becker, A. Zolan, K. Anderson, D. Olis and E. Elgqvist, "Computational Framework for Behind-The-Meter DER Techno-Economic Modeling and Optimization - REopt Lite," *working paper (available arXiv.org),* 2020.

[8] O. Ogunmodede, K. Anderson, D. Cutler and A. Newman, "Optimizing design and dispatch of a renewable energy system," *working paper,* 2020.

[9] R. Lite API, "REopt Lite Open Source Code (https://github.com/NREL/REopt_Lite_API/blob/master/resilience_stats/outage_simulator_LF.py)," National Renewable Energy Laboratory, 2020.

[10] OpenEI, "U.S. URDB, https://openei.org/apps/USURDB/rate/view/5ed6b8535457a3357add15ab".

[11] S. Mishra, "Case Study Scenario I results (https://reopt.nrel.gov/tool/results/8f358e7f-4669-4062-b00e-c8bfed9f710d)".

[12] S. Mishra, "Case Study Scenario II results (https://reopt.nrel.gov/tool/results/1eace892-b0cc-4f60-8192-194b030467a6)".

[13] NREL, "REopt Lite Webtool".